\documentclass[12pt,aps,prd,showpacs,amsmath,amssymb]{revtex4}
\usepackage{rotating}
\input epsf
\newcommand{\doslash}[1]{#1{\!\!\!\slash}}
\begin{document}
\title{Chiral symmetry-breaking corrections to strong decays of $D_{s0}^{*}(2317)$ and $D_{s1}^{'}(2460)$ in HH$\chi$PT}
\author{Jin-Yun Wu, Yong-Lu Liu, Jian-Rong Zhang, and Ming-Qiu Huang}
\affiliation{ College of Science, National University of Defense Technology, Hunan, 410073, China}
\date{\today}
\begin{abstract}
The strong decays of two narrow mesons $D_{s0}^{*}(2317)$ and
$D_{s1}^{'}(2460)$ are studied within the framework of heavy hadron chiral
perturbation theory. Up to next-to-leading order in $1/\Lambda_{\chi}$, by a fit to
the experimental widths of their nonstrange partners,
the chiral symmetry-breaking coupling constants are extracted. The single-pion
decay widths are estimated to be $\Gamma(D_{s0}^{*}(2317)\to
D_{s}^{+}\pi^{0})=9.2\pm2.3$ KeV and $\Gamma(D_{s1}^{'}(2460)\to
D_{s}^{*+}\pi^{0})=9.0\pm2.1$ KeV, respectively, which are consistent with the
experimental constraints and comparable with other theoretical predictions.
The numerical analysis shows that chiral-symmetry corrections to the decay widths are significant.
Applications and predictions for the corresponding beauty mesons are also
provided.
\end{abstract}
 \maketitle

\section{Introduction}
\label{intro}

In the last decade, many open charm or hidden
charm heavy mesons were discovered, which contribute to  the revival of hadron
spectroscopy. Two outstanding mesons among them are the narrow mesons
$D_{s0}^{*}(2317)$ and $D_{s1}^{'}(2460)$, observed  in the final states
$D_{s}^{+}\pi^{0}$ and $D_{s}^{*+}\pi^{0}$  \cite{exp1}, which are
naturally assigned with the quantum numbers $J^{p}=0^{+}, 1^{+}$. The puzzle is that their measured masses and widths do not match the
predictions from potential-based quark models \cite{Godfrey1991},
unexpectedly, i.e. they lie below $DK$ and $D^{*}K$ thresholds respectively
and their widths are extremely narrow. Since their discoveries, there have
been lots of experimental investigations\cite{Belle,FOCUS,BABAR}. Meanwhile, many theoretical papers are dedicated
to the understanding of their underlying structures. Proposed schemes include the conventional
$c\bar{s}$ $(0^{+}, 1^{+})$ chiral partners of the $(D_{s}, D_{s}^{*})$
doublet in HQET
\cite{Colangelo2012,Mehen2004,Bardeen2003,Fayyazuddin2004,Wei2006,Lu2006,Godfrey2003,Liu2006,Colangelo2003,Ishida2004},
$DK$ molecules
\cite{Barnes2003,ValeryLyubovitskij2007,Guo2014,Guo2008,Guo2013},
four-quark states \cite{Cheng2003,Nielsen2006,Browder2004}, $D\pi$ atoms
\cite{Szczepaniak2003}, $c\bar{s}$-$c\bar{s}q\bar{q}$ admixtures
\cite{Terasaki2003}, and admixture of $c\bar{s}$ and $DK$-molecule (for
$D_{s0}^{*}(2317)$) \cite{Mohler2013}.

Quantities, which have
different values in different interpretations, would be useful to
distinguish them, such as the decay modes.
However, as masses of these two states are lower than the $DK$ and $D^{*}K$
thresholds respectively, the potentially dominant s-wave decay modes
$D_{s0}^{*}(2317) \rightarrow DK$ and $D_{s1}^{'}(2460) \rightarrow D^{*}K$
 are kinematically forbidden. Therefore, the isospin violating strong decays and radiative decays are the
promising quantities.
In literatures, many discussions of their strong and radiative decays,
and the decays into them  from the beauty mesons have been presented
\cite{Mehen2004,Bardeen2003,Fayyazuddin2004,Wei2006,Lu2006,Godfrey2003,Liu2006,Colangelo2003,Ishida2004,ValeryLyubovitskij2007,Guo2014,Guo2008,Guo2013,Cheng2003,Nielsen2006,Mohler2013,Segovia2012}.
Moreover, the branching ratios of their strong and radiative decays were measured quite
accurately by Belle Collaboration \cite{Belle} and BABAR Collaboration
\cite{BABAR}. Nevertheless, the single-pion strong decay widths, one of the most important quantities,
differ significantly from various approaches.
It can be concluded that they are several tens of \mbox{KeV} in the $c\bar{s}$ scenario for the small
$\eta-\pi^{0}$ mixing angle $\sim10^{-2}$ (see e.g. results in Ref. \cite{Wei2006}),
while near one hundred \mbox{KeV} in other scenarios due to
additional direct strong isospin-violating transitions
(see e.g. discussions in Ref. \cite{ValeryLyubovitskij2007}).
However, a direct experimental judgement of this still needs to be
found. Furthermore, decay widths of
their observed non-strange partners cannot be well fitted by just leading
order contributions, as can be seen from the discussions in Ref. \cite{Colangelo2012,Mehen2004}.
To decipher this discrepancy, a more
careful calculation of their strong decay widths will be very helpful.

\begin{figure}[bp]
\centering
\begin{minipage}[t]{0.45\linewidth}
    \centering
    \includegraphics[width=5cm,height=2.5cm]{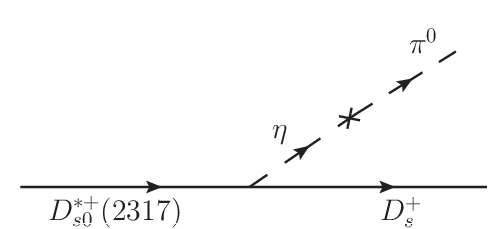}
    \parbox{3.5cm}{\small \hspace{0cm}(a) }
\end{minipage}
\hspace{2ex}
\begin{minipage}[t]{0.45\linewidth}
    \centering
    \includegraphics[width=5cm,height=2.5cm]{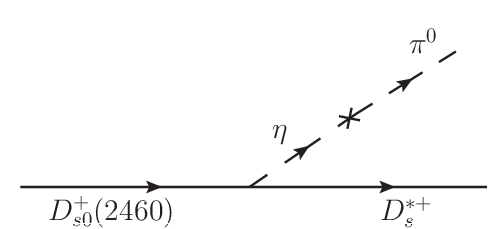}
    \parbox{3.5cm}{\small \hspace{0cm}(b) }
\end{minipage}
\caption{(a) $D_{s0}^{*+}(2317)\rightarrow D_{s}^{+}+\eta\rightarrow D_{s}^{+}+\pi^{0}$,\;
  (b) $D_{s1}^{'+}(2460)\rightarrow D_{s}^{*+}+\eta\rightarrow D_{s}^{*+}+\pi^{0}$.}
  \label{2317decayfig}
\end{figure}

In this work, we assume these two states as the $c\bar{s}$ $(0^{+}, 1^{+})$ chiral
partners of the  $(D_{s}, D_{s}^{*})$ $H$ doublet in HQET and calculate the single-pion strong decays of $D_{s0}^{*}(2317)$
and $D_{s1}^{'}(2460)$ by taking into account the chiral symmetry-breaking corrections within the framework of heavy hadron chiral perturbation theory ($HH\chi PT$)
\cite{chiraltheory}. The method is a combination of HQET and chiral perturbation theory.
The decays occur through two steps: $D_{s0}^{*}(2317) \rightarrow D_{s}+\eta \rightarrow
D_{s}+\pi^{0}$ and $D_{s1}^{'}(2460) \rightarrow D_{s}^{*}+\eta \rightarrow
D_{s}^{*}+\pi^{0}$, shown in Fig. \ref{2317decayfig}. As is known, the mass of $s$ quark is much larger than that of $u$ and $d$.
Therefore, the chiral symmetry-breaking corrections are expected to be significant.

To calculate the chiral-symmetry breaking corrections in $HH\chi PT$, a large amount of
unknown coupling constants need to be determined with the experimentally measured decay rates of
the $S$ doublet mesons listed in Table \ref{Dmesonsdata}. However, before 2015, there had existed a puzzle on
the charged and neutral $0^{+}$ states $D_{0}^{*\pm}$ and $D_{0}^{*0}$ ever since they were discovered in 2004
\cite{Belle2004,FOCUSD00}, which are predicted to be degenerated in both masses and decay widths in
the quark model \cite{Godfrey1991}.
In other words, though their measured widths were degenerate within the errors, the measured masses were severely splitted, which would
result in very different decay rates in $HH\chi PT$ because of the different phase spaces. But it is exciting that, in 2015, the new experiments on $D_{0}^{*\pm}$
done by LHCb collaboration \cite{LHCb15X,LHCb15Y} greatly suppressed the gap between the masses of $D_{0}^{*\pm}$ and $D_{0}^{*0}$:
\begin{equation}\label{massdiscimp}
  M_{D_{0}^{*\pm}}-M_{D_{0}^{*0}}\sim90 \text{MeV}\text{(before 2015)}
  \rightarrow \sim30 \text{MeV}\text{(after 2015)},
\end{equation}
as can be seen from Table \ref{D00data}. This suppression helps to determine the chiral symmetry-breaking coupling constants and finally predict more accurately the decay rates of $D_{s0}^{*}(2317)$ and $D_{s1}^{'}(2460)$.

\begin{table}
\caption{\label{D00data} Renewed experimental values of the masses and widths of $D_{0}^{*\pm}$ \cite{PDG2016}.}
\begin{tabular}{cc l}
\hline\noalign{\smallskip}
\hline\noalign{\smallskip}
$Mass(\text{MeV})$ & $\Gamma(\text{MeV})$ & Comment\\
\noalign{\smallskip}\hline\noalign{\smallskip}
$2351\pm7$ &  $230\pm 17$ & PDG average (2016)\cite{PDG2016} \\
$2360\pm15\pm30$  & $255\pm26\pm51$ & LHCb $B^{0}\rightarrow\bar{D}^{0}K^{+}\pi^{-}$ (2015)\cite{LHCb15X}\\
$2349\pm6\pm4$  & $217\pm13\pm13$ & LHCb $B^{0}\rightarrow\bar{D}^{0}\pi^{+}\pi^{-}$ (2015)\cite{LHCb15Y}\\
$2403\pm14\pm35$ & $283\pm24\pm34$ & FOCUS $\gamma A$ (2004)\cite{FOCUSD00} \\
\noalign{\smallskip}\hline
\hline\noalign{\smallskip}
\end{tabular}
\end{table}

\begin{table}
\caption{\label{Dmesonsdata} Experimentally measured masses and widths of the observed $S$
doublet heavy-light mesons and observed single-pion strong decay modes (SPSDMs). All the results are from the
PDG \cite{PDG2016}, and the quoted bounds are at 95\% CL.}
\begin{tabular}{ccccc}
\hline\hline\noalign{\smallskip}
            & $J^{P}$ &$Mass(\text{MeV})$ & $\Gamma(\text{MeV})$ & observed SPSM\\
\noalign{\smallskip}\hline\noalign{\smallskip}
$D^{*0}_{0}(2400)$ & $0^{+}$ &$2318\pm29$ & $267\pm40$ & $D^{+}\pi^{-}$\\
$D^{*\pm}_{0}(2400)$ & $0^{+}$ & $2351\pm7$ & $230\pm17$ & $D^{0}\pi^{+}$\\
$D^{'0}_{1}(2430)$ & $1^{+}$ &$2427\pm26\pm25$ & $384\pm^{107}_{75}\pm74$& $D^{*+}\pi^{-}$ \\
$D^{*}_{s0}(2317)$ & $0^{+}$ &$2317.7\pm0.6$ & $<3.8$ & $D^{+}_{s}\pi^{0}$, $D^{*+}_{s}\pi^{0}$\\
$D^{'}_{s1}(2460)$ & $1^{+}$ &$2459.5\pm0.6$ & $<3.5$ & $D^{*+}_{s}\pi^{0}$, $D^{+}_{s}\pi^{0}$\\
\noalign{\smallskip}\hline\hline
\end{tabular}
\end{table}

This paper is organized as follows. In Section \ref{sec2}, we incorporate the
doublets into the effective heavy hadron chiral Lagrangian, which is
written out to terms of next-to-leading order in $1/\Lambda_{\chi}$. In
Section \ref{sec3}, we discuss single-pion strong decays of charmed heavy mesons
and the corresponding beauty ones in the heavy quark spin-flavor symmetry.
Numerical calculation and the results are discussed in Section \ref{sec4},
including a brief summary.

\section{The chiral Lagrangian}
\label{sec2}

 The strong decays of excited heavy-light mesons involve the
emission of soft pions and kaons, and hence it is useful to analyze these
interactions with the chiral perturbation theory \cite{chiraltheory}.
The octet of light pseudoscalar mesons is introduced through the definition
$\Sigma=\xi^{2}={\mbox{exp}}(2i\mathcal{M}/f_{\pi})$, where
\begin{eqnarray}
  \mathcal{M} &=& \pi^{i}\lambda^{i}=\left(
  \begin{array}{ccc}
  \frac{1}{\sqrt{2}}\pi^{0}+\frac{1}{\sqrt{6}}\eta & \pi^{+} & K^{+} \\
  \pi^{-} & \frac{-1}{\sqrt{2}}\pi^{0}+\frac{1}{\sqrt{6}}\eta & K^{0} \\
  K^{-} & \bar{K}^{0} & -\frac{\sqrt{2}}{\sqrt{3}}\eta \\
  \end{array}
  \right).
\end{eqnarray}

The heavy-light mesons are customarily cataloged by the total
angular momentum of the light degrees of freedom $s_{l}^{p}$ ($p$ denotes
the parity), which is a good quantum number because of heavy quark spin
symmetry in the heavy quark limit $m_{Q}\rightarrow \infty$. In this paper, only two doublets,
$H$ doublet ($0^{-}, 1^{-}$) and $S$ doublet ($0^{+}, 1^{+}$), corresponding to $s_{l}^{p}=1/2^{-}, 1/2^{+}$
are discussed, which can be respectively represented by the superfields
$H_{a}=\frac{1+\doslash{v}}{2}[P^{*}_{a\mu}\gamma^{\mu}-P_{a}\gamma_{5}]$
($a=u,d,s$, a light flavor index), where $P^{*}_{a\mu}$ and $P_{a}$
annihilate the vector and pseudoscalar mesons, and
$S_{a}=\frac{1+\doslash{v}
}{2}[P^{'}_{1a\mu}\gamma^{\mu}\gamma_{5}-P_{0a}^{*}]$ for the axial-vector
$P^{'}_{1a\mu}$ and scalar $P_{0a}^{*}$ mesons.

Considering heavy quark spin-flavor symmetry and light quark chiral symmetry,
an effective Lagrangian responsible for the strong decay
$S\rightarrow HM$ ($M$ is a light pseudoscalar meson) can be written with
these superfields. The leading order contribution in $1/\Lambda_{\chi}$ and
$1/m_{Q}$ is
\begin{equation}\label{Lag-1}
  \mathcal{L}_{mix}=hTr[\bar{H}_{b}S_{a}\doslash{\mathcal{A}}_{ab}\gamma_{5}]+h.c..
\end{equation}
According to Refs.\cite{chiral-L,Fajfer2006}, the corresponding chiral
symmetry breaking corrections to the Lagrangian Eq. (\ref{Lag-1}) to next-to-leading
order in $1/\Lambda_{\chi}$ read
\begin{eqnarray}
\mathcal{L}_{mix}^{sb}&&= 1/\Lambda_{\chi}\{\kappa_{1}Tr[(\bar{H}S\doslash{\mathcal{A}}\gamma_{5})_{ab}(m^{\xi}_{q})_{ba}]
+\kappa_{2}Tr[(\bar{H}S\doslash{\mathcal{A}}\gamma_{5})_{aa}(m^{\xi}_{q})_{bb}]
\nonumber\\&&{}+\kappa_{3}Tr[\bar{H}_{a}S_{a}\doslash{\mathcal{A}}_{bc}\gamma_{5}(m^{\xi}_{q})_{cb}]
+\kappa_{4}Tr[\bar{H}_{c}S_{a}\doslash{\mathcal{A}}_{bc}\gamma_{5}(m^{\xi}_{q})_{ab}]
\nonumber\\&&{}+\kappa_{5}Tr[\bar{H}_{a}S_{b}iv\cdot\mathcal{D}_{bc}\doslash{\mathcal{A}}_{ca}\gamma_{5}]
+\kappa_{6}Tr[\bar{H}_{a}S_{b}i\doslash{\mathcal{D}}_{bc}v\cdot\mathcal{A}_{ca}\gamma_{5}]
\}+h.c..
\end{eqnarray}
Meanwhile, the effective Lagrangian responsible for $\eta-\pi^{0}$ mixing,
through which  the pionic decays of $D_{s0}^{*}(2317)$ and
$D_{s1}^{'}(2460)$ occur, can be described by the isospin violating piece
in the chiral Lagrangian
\begin{eqnarray}
\mathcal{L}_{\eta-\pi^{0}}&=&
\frac{m^{2}_{\pi}f^{2}_{\pi}}{4(m_{u}+m_{d})}Tr[m_{q}^{\dagger}\Sigma+\Sigma^{\dagger}m_{q}]\nonumber\\
&=& \frac{m^{2}_{\pi}(m_{u}-m_{d})}{\sqrt{3}(m_{u}+m_{d})}\pi^{0}\eta+\cdots.
\end{eqnarray}

Herein, $\bar{H}_{a}=\gamma^{0}H^{\dagger}_{a}\gamma^{0}$,
and $\mathcal{D}_{ab}^{\mu}=\delta_{ab}\partial^{\mu}-\mathcal{V}_{ab}^{\mu}$. In the expressions,
$\mathcal{V}_{\mu}=1/2(\xi^{\dagger}\partial_{\mu}\xi+\xi\partial_{\mu}\xi^{\dagger})$
and
$\mathcal{A}_{\mu}=i/2(\xi^{\dagger}\partial_{\mu}\xi-\xi\partial_{\mu}\xi^{\dagger})$
are the light meson vector and axial currents, containing an
even number and an odd number of pseudoscalar fields, respectively.
$\mathcal{D}_{ab}^{\mu}\mathcal{A}_{bc}^{\nu}=\partial^{\mu}\mathcal{A}_{ac}^{\nu}
+[\mathcal{V}^{\mu},\mathcal{A}^{\nu}]_{ac}$.
And the chiral symmetry-breaking scale $\Lambda_{\chi}$ is set to be $\Lambda_{\chi}=1$ GeV. The
$3\times3$ mass matrix is $m_{q}=diag(m_{u},m_{d},m_{s})$, and $m_{q}^{\xi}=\xi
m_{q}\xi+\xi^{\dagger}m_{q}\xi^{\dagger}$.

Note that a full calculation of the strong decays should also contain, in
addition to the chiral symmetry-breaking corrections, the heavy quark
symmetry breaking corrections in $1/m_{Q}$. However, if the $1/m_{Q}$ corrections are
also included, the number of free parameters to be determined will be too larger compared with the number of
experimentally measured decay rates of the $S$ doublet heavy mesons, thus heavily weakens
the effectiveness of the $\chi^2$ fitting.
Moreover, the lattice QCD studies \cite{McNeile2004,Abada2004}
of the strong couplings of heavy mesons indicate that these  $1/m_{Q}$ corrections
seem not to be significant but pointed out the importance of controlling chiral corrections.
And based on these lattice researches, chiral loop corrections to strong decays of non-strange charmed mesons in
$S$ doublet
have been studied in the Ref. \cite{Fajfer2006}, obtaining pretty good results.
Therefore, we just concentrate on the chiral symmetry-breaking corrections,
while ignore the heavy quark symmetry corrections in the calculation.

\section{Single-pion decay of excited heavy mesons}

\label{sec3}

\begin{figure}[bp]
\centering
\begin{minipage}[t]{0.45\linewidth}
    \centering
    \includegraphics[width=5cm,height=2.5cm]{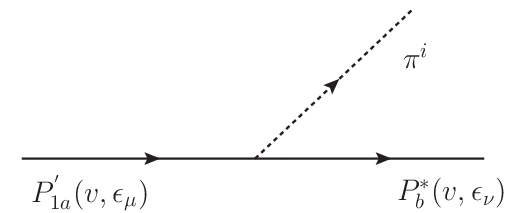}
    \parbox{3.5cm}{\small \hspace{0cm}(a) }
\end{minipage}
\hspace{2ex}   
\begin{minipage}[t]{0.45\linewidth}
    \centering
    \includegraphics[width=5cm,height=2.5cm]{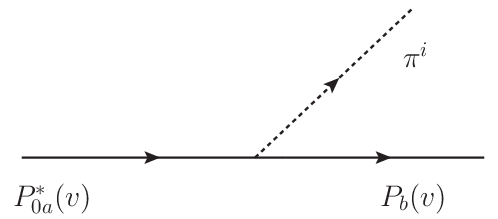}
    \parbox{3.5cm}{\small \hspace{0cm}(b) }
\end{minipage}
\caption{(a) $P^{'}_{1a}\rightarrow P^{*}_{b}\pi^{i}$,\;\;
  (b) $P^{*}_{0a}\rightarrow P_{b}\pi^{i}$.}\label{decayfig}
\end{figure}

Using the Lagrangian given in Section \ref{sec2}, the formulae of the
single-pion decays $S_{a}\rightarrow H_{b}\pi^{i}$ ($a,b=u,d,s$ and
$i=1,2,\cdots,8$), shown in Fig. \ref{decayfig}, are
\begin{subequations}
\begin{eqnarray}
\Gamma(P^{'}_{1a}\rightarrow P^{*}_{b}\pi^{i}) &=& \frac{1}{8\pi}\frac{M_{P^{*}_{b}}}
{M_{P^{'}_{1a}}}E^{2}_{\pi^{i}}|\vec {P}_{\pi^{i}}|\theta^{2}_{ab}F^{i2}_{ab}, \\
\Gamma(P^{*}_{0a}\rightarrow P_{b}\pi^{i}) &=&
\frac{1}{8\pi}\frac{M_{P_{b}}}{M_{P^{*}_{0a}}}E^{2}_{\pi^{i}}|\vec
{P}_{\pi^{i}}|\theta^{2}_{ab}F_{ab}^{i2}\;,
\end{eqnarray}\label{generaldecay}
\end{subequations}
where the decay amplitudes $F^{i}_{ab}$ read
\begin{eqnarray}
  F^{i}_{ab} &&= \frac{2h}{f_{\pi}}\lambda^{i}_{ab}
  +\frac{4\kappa_{1}}{\Lambda_{\chi}f_{\pi}}\lambda_{ac}^{i}(m_{q})_{cb}
+\frac{4\kappa_{2}}{\Lambda_{\chi}f_{\pi}}\lambda_{ab}^{i}(m_{q})_{cc}
  \nonumber\\&&{}+\frac{4\kappa_{3}}{\Lambda_{\chi}f_{\pi}}\lambda_{cd}^{i}(m_{q})_{dc}\delta_{ab}
  +\frac{4\kappa_{4}}{\Lambda_{\chi}f_{\pi}}\lambda_{cb}^{i}(m_{q})_{ac}
  -\frac{2\kappa_{5}}{\Lambda_{\chi}f_{\pi}}\lambda_{ab}^{i}E_{\pi^{i}}-\frac{2\kappa_{6}}{\Lambda_{\chi}f_{\pi}}\lambda_{ab}^{i}E_{\pi^{i}}\;.
\end{eqnarray}
In the expressions, $\theta_{ab}=\theta$ for $ab=33$ (while $1$ for other $ab$s),
$\kappa^{'}_{5}=\kappa_{5}+\kappa_{6}$,  and $\lambda^{i}$ is the
corresponding coefficient matrix of $\pi^{i}$ in the definition of
$\mathcal{M}=\pi^{i}\lambda^{i}$.
The $\eta-\pi^{0}$ mixing angle is
$\displaystyle{\theta=\frac{\sqrt{3}}{4}\frac{m_{d}-m_{u}}{m_{s}-(m_{u}+m_{d})/2}}$,
 accounting for the isospin violation. From this formula, as mentioned in Ref. \cite{Stewart1998}, the following transformations of the parameters are helpful:
 \begin{description}\label{transformations}
   \item[(i)] As $\kappa_{2}$ can be absorbed into the definition of $h$, we set
   $h^{'}=h+\frac{2(m_{u}+m_{d}+m_{s})}{\Lambda_{\chi}}$.
   \item[(ii)] As $\kappa_{5}$ and $\kappa_{6}$ always enter in a fixed combination, they are properly represented by a united parameter
    $\kappa_{5}^{'}=\kappa_{5}+\kappa_{6}$.
   \item[(iii)] For $\kappa_{1}$ and $\kappa_{4}$, if we define $\kappa_{1}^{'}=\frac{\kappa_{1}+\kappa_{4}}{2}$
   and $\kappa_{4}^{'}=\frac{\kappa_{1}-\kappa_{4}}{2}$, they will be distinguishable that $\kappa_{1}^{'}$ concerns only the isospin conserving contributions, while $\kappa_{4}^{'}$ involves only the isospin violating contributions.
 \end{description}

Finally, we rewrite the decay amplitudes as follows:
  \begin{eqnarray}
    \text{for $c\bar{q}$ mesons},\nonumber\\
    F^{i}_{ab}&&= \frac{2h^{'}}{f_{\pi}}
  +\frac{4\kappa_{1^{'}}}{\Lambda_{\chi}f_{\pi}}(m_{u}+m_{d})
  +\frac{4\kappa_{4^{'}}}{\Lambda_{\chi}f_{\pi}}(m_{d}-m_{u})g_{ab}
  \nonumber\\&&{}-\frac{4\kappa_{3}}{\Lambda_{\chi}f_{\pi}}(m_{d}-m_{u})\delta_{ab}
  -\frac{2\kappa_{5}^{'}}{\Lambda_{\chi}f_{\pi}}E_{\pi^{i}}, \label{cqamplitude}\\
  \text{while for $c\bar{s}$ mesons}, \nonumber\\
    F^{i}_{ab}&&= \frac{2h^{'}}{f_{\pi}}
  +\frac{8\kappa_{1^{'}}}{\Lambda_{\chi}f_{\pi}}m_{s}
  +\frac{2\kappa_{3}}{\Lambda_{\chi}f_{\pi}}(2m_{s}-m_{d}-m_{u})
  -\frac{2\kappa_{5}^{'}}{\Lambda_{\chi}f_{\pi}}E_{\pi^{i}}, \label{csamplitude}
  \end{eqnarray}
where $g_{ab}$ is an auxiliary sign function which satisfies $g_{ab}=1$ for $a<b$ and $g_{ab}=-1$ for $a\geq b$. Thus we can see that by these transformations the seven undetermined parameters are cut down to five in the $c\bar{q}$ case and to four in the $c\bar{s}$ case where the isospin violating terms ($\kappa_{4}^{'}$) no longer appear. Furthermore, it is remarkable that the $\kappa_{3}$ terms, which give the isospin violating contributions in the $c\bar{q}$ case when the final pion state is $\pi^{0}$, become important in the $c\bar{s}$ case where they contribute to the isospin conserving transitions. Therefore, we don't
choose to neglect the isospin violating effects in our calculation.

The experimentally observed excited heavy-light mesons of $S$ doublet are $D^{*0}_{0}$ $(2400)$,
$D^{*\pm}_{0}(2400)$,  $D^{'0}_{1}(2430)$, $D^{*}_{s0}(2317)$ and
$D^{'}_{s1}(2460)$. Their measured masses and widths, as well as the observed single-pion strong decay modes (SPSDMs), are listed in Table \ref{Dmesonsdata}. We can see from the data that:
\begin{description}
  \item[(i)\;]   Though the masses of $D_{0}^{*0}$ and $D_{0}^{*+}$ are still decoupled, the discrepancy has been greatly shrinked by the new experiments in 2015. Thus it will do good for the
      determination of the $\kappa_{i}$s.
  \item[(ii)] Though the two SPSDMs $D_{s0}^{*}D_{s}^{*+}\pi^{0}$ and $D_{s1}^{'}D_{s}^{+}\pi^{0}$ are also listed in the table, we don't take them into account since they are actually negligible as we can see from Ref. \cite{PDG2016}:
      \begin{equation}
        \frac{\Gamma(D_{s0}^{*}(2317)\rightarrow D_{s}^{*+}\pi^{0})}{\Gamma(D_{s0}^{*}(2317)\rightarrow D_{s}^{+}\pi^{0})}< 0.11,
      \end{equation}
      and
      \begin{eqnarray}
        &&\frac{\Gamma(D_{s1}^{'}(2460)\rightarrow D_{s}^{+}\pi^{0})}{\Gamma(D_{s1}^{'}(2460)
        \rightarrow D_{s}^{*+}\pi^{0})+\Gamma(D_{s1}^{'}(2460)\rightarrow D_{s0}^{*+}(2317)\gamma)}< 0.042,\;\nonumber\\
        &&\frac{\Gamma(D_{s1}^{'}(2460)\rightarrow D_{s0}^{*+}(2317)\gamma)}
        {\Gamma(D_{s1}^{'}(2460)\rightarrow D_{s}^{*+}\pi^{0})+\Gamma(D_{s1}^{'}(2460)\rightarrow D_{s0}^{*+}(2317)\gamma)}< 0.25.
      \end{eqnarray}
\end{description}

Additionally, it should be aware that the two $1^{+}$ states $D^{'0}_{1}(2430)$ and $D^{0}_{1}(2420)$, corresponding to $s_{l}^{p}=1/2^{+}$ and $s_{l}^{p}=3/2^{+}$ respectively,
may mix with each other.
However, the experimental measurement of Belle Collaboration \cite{Belle2004} says that the
mixing angle is $\omega=-0.10\pm0.03\pm0.02\pm0.02$,
suggesting that such a mixing can be safely neglected \cite{Colangelo2012} and $D^{'0}_{1}(2430)$ can be regarded as $1^{+}(s_{l}=\frac{1}{2})$ state.
 Thus we don't include the mixing of these two axial states in this work.

In the bottom sector, no candidate of the $S$ doublet bottom partners has ever been observed. However,
in the heavy quark limit, the heavy quark flavor symmetry guarantees that the chiral symmetry-breaking
coupling constants between the bottom doublets are the same as the coupling constants between the corresponding charm doublets. Thus with $h$ and $\kappa_{i}$s determined by the experimental data in Table \ref{Dmesonsdata}, we obtain the decay rates of SPSDMs of the mesons in bottom $S$ doublets by taking use of their predicted masses listed in Table \ref{Bmesonsdata} in the same framework of $HH\chi PT$ \cite{Colangelo2012}.

\begin{table}[ht]
\caption{\label{Bmesonsdata} Predicted masses \cite{Colangelo2012} of experimentally unobserved $S$
doublet bottom partners under the same theoretical framework of this work and the suggested SPDMs.}
\begin{tabular}{cccc}
\hline\hline\noalign{\smallskip}
                   & $J^{P}$ &$Mass(\mbox{MeV})$ & Suggested SPDMs\\
\noalign{\smallskip}\hline\noalign{\smallskip}
$B^{*0}_{0}$ & $0^{+}$ & $5708.2\pm22.5$ & $B^{+}\pi^{-}$, $B^{0}\pi^{0}$ \\
$B^{*\pm}_{0}$ & $0^{+}$ & $5708.2\pm22.5$  & $B^{0}\pi^{+}$, $B^{0}\pi^{\pm}$\\
$B^{'0}_{1}$ & $1^{+}$ & $5753.3\pm31.1$ & $B^{*+}\pi^{-}$, $B^{*0}\pi^{0}$\\
$B^{*}_{s0}$ & $0^{+}$ & $5706.6\pm1.2$ & $B^{+}_{s}\pi^{0}$\\
$B^{'}_{s1}$ & $1^{+}$ & $5765.6\pm1.2$ & $B^{*+}_{s}\pi^{0}$\\
\noalign{\smallskip}\hline\hline
\end{tabular}
\end{table}

Based on the formulae Eqs. (\ref{generaldecay}) and (\ref{cqamplitude}), decay widths of the exclusive transitions from  an observed $c\bar{q}$ $S$ doublet meson to a $c\bar{q}$ $H$ doublet
 meson and a single pion (shown in Fig. \ref{decayfig}) are obtained as

\begin{subequations}\label{decayrates}
\begin{eqnarray}
\Gamma(D^{*0}_{0}\rightarrow D^{+}\pi^{-}) &&= \frac{1}{8\pi}
\frac{M_{D^{+}}}{M_{D^{*0}_{0}}}E^{2}_{\pi^{-}}|\vec
{P}_{\pi^{-}}|[\frac{2h^{'}}{f_{\pi}}+\frac{4(m_{d}+m_{u})}{\Lambda_{\chi}f_{\pi}}\kappa^{'}_{1}
\nonumber\\&&{}
+\frac{4(m_{d}-m_{u})}{\Lambda_{\chi}f_{\pi}}\kappa^{'}_{4}
   -\frac{2E_{\pi^{-}}}{\Lambda_{\chi}f_{\pi}}\kappa_{5}^{'}]^{2}\;,\\
\Gamma(D^{*+}_{0}\rightarrow D^{0}\pi^{+}) &&= \frac{1}{8\pi}
\frac{M_{D^{0}}}{M_{D^{*+}_{0}}}E^{2}_{\pi^{+}}|\vec
{P}_{\pi^{+}}|[\frac{2h^{'}}{f_{\pi}}+\frac{4(m_{d}+m_{u})}{\Lambda_{\chi}f_{\pi}}\kappa^{'}_{1}
\nonumber\\&&{} -\frac{4(m_{d}-m_{u})}{\Lambda_{\chi}f_{\pi}}\kappa^{'}_{4}
     -\frac{2E_{\pi^{+}}}{\Lambda_{\chi}f_{\pi}}\kappa_{5}^{'}]^{2}\;,\\
\Gamma(D^{'0}_{1}\rightarrow D^{*+}\pi^{-}) &&= \frac{1}{8\pi}
\frac{M_{D^{*+}}}{M_{D^{'0}_{1}}}E^{2}_{\pi^{-}}|\vec
{P}_{\pi^{-}}|[\frac{2h^{'}}{f_{\pi}}+\frac{4(m_{d}+m_{u})}{\Lambda_{\chi}f_{\pi}}\kappa^{'}_{1}
\nonumber\\&&{}
 +\frac{4(m_{d}-m_{u})}{\Lambda_{\chi}f_{\pi}}\kappa^{'}_{4}
      -\frac{2E_{\pi^{-}}}{\Lambda_{\chi}f_{\pi}}\kappa_{5}^{'}]^{2}\;,\\
\Gamma(D^{*0}_{0}\rightarrow D^{0}\pi^{0}) &&= \frac{1}{16\pi}
\frac{M_{D^{0}}}{M_{D^{*0}_{0}}}E^{2}_{\pi^{0}}|\vec
{P}_{\pi^{0}}|[\frac{2h^{'}}{f_{\pi}}+\frac{4(m_{d}+m_{u})}{\Lambda_{\chi}f_{\pi}}\kappa^{'}_{1}
 \nonumber\\&&{}-\frac{4(m_{d}-m_{u})}{\Lambda_{\chi}f_{\pi}}\kappa^{'}_{4}
   -\frac{4(m_{d}-m_{u})}{\Lambda_{\chi}f_{\pi}}\kappa_{3}
  -\frac{2E_{\pi^{0}}}{\Lambda_{\chi}f_{\pi}}\kappa_{5}^{'}]^{2} \;,\\
\Gamma(D^{*+}_{0}\rightarrow D^{+}\pi^{0}) &&= \frac{1}{16\pi}
\frac{M_{D^{+}}}{M_{D^{*+}_{0}}}E^{2}_{\pi^{0}}|\vec
{P}_{\pi^{0}}|[\frac{2h^{'}}{f_{\pi}}+\frac{4(m_{d}+m_{u})}{\Lambda_{\chi}f_{\pi}}\kappa^{'}_{1}
 \nonumber\\&&{}-\frac{4(m_{d}-m_{u})}{\Lambda_{\chi}f_{\pi}}\kappa^{'}_{4}
   -\frac{4(m_{d}-m_{u})}{\Lambda_{\chi}f_{\pi}}\kappa_{3}
   -\frac{2E_{\pi^{0}}}{\Lambda_{\chi}f_{\pi}}\kappa_{5}^{'}]^{2} \;,\\
\Gamma(D^{'0}_{1}\rightarrow D^{*0}\pi^{0}) &&= \frac{1}{16\pi}
\frac{M_{D^{*0}}}{M_{D^{'0}_{1}}}E^{2}_{\pi^{0}}|\vec
{P}_{\pi^{0}}|[\frac{2h^{'}}{f_{\pi}}+\frac{4(m_{d}+m_{u})}{\Lambda_{\chi}f_{\pi}}\kappa^{'}_{1}
 \nonumber\\&&{}-\frac{4(m_{d}-m_{u})}{\Lambda_{\chi}f_{\pi}}\kappa^{'}_{4}
  -\frac{4(m_{d}-m_{u})}{\Lambda_{\chi}f_{\pi}}\kappa_{3}
   -\frac{2E_{\pi^{0}}}{\Lambda_{\chi}f_{\pi}}\kappa_{5}^{'}]^{2}\;.
\end{eqnarray}
\end{subequations}

As to $D_{s0}^{*}(2317)$ and $D_{s1}^{'}(2460)$, their decays to the corresponding
$H$ doublet mesons and the single pion state $\pi^{0}$ occur through the intermediate $\eta$ meson, as shown in Fig. \ref{2317decayfig}. The decay widths are

\begin{subequations}\label{23142460decayrates}
\begin{eqnarray}
&\Gamma(D^{*+}_{s0}&(2317)\rightarrow D^{+}_{s}\eta\rightarrow
D^{+}_{s}\pi^{0})  = \frac{1}{12\pi}
\frac{M_{D^{+}_{s}}}{M_{D^{*+}_{s0}}}E^{2}_{\pi^{0}}
|\vec{P}_{\pi^{0}}|\theta^{2}\nonumber\\&&{}
\times[\frac{2h^{'}}{f_{\pi}}+\frac{8m_{s}}{\Lambda_{\chi}f_{\pi}}\kappa^{'}_{1}+
+\frac{2(2m_{s}-m_{u}-m_{d})}{\Lambda_{\chi}f_{\pi}}\kappa_{3}
   -\frac{2E_{\pi^{0}}}{\Lambda_{\chi}f_{\pi}}\kappa_{5}^{'}]^{2}\;, \\
&\Gamma(D^{'+}_{s1}&(2460)\rightarrow D^{*+}_{s}\eta\rightarrow
D^{*+}_{s}\pi^{0})  = \frac{1}{12\pi}
\frac{M_{D^{*+}_{s}}}{M_{D^{'+}_{s1}}}E^{2}_{\pi^{0}}|\vec{P}_{\pi^{0}}|\theta^{2}\nonumber\\&&{}
\times[\frac{2h^{'}}{f_{\pi}}+\frac{8m_{s}}{\Lambda_{\chi}f_{\pi}}\kappa^{'}_{1}+
 +\frac{2(2m_{s}-m_{u}-m_{d})}{\Lambda_{\chi}f_{\pi}}\kappa_{3}
    -\frac{2E_{\pi^{0}}}{\Lambda_{\chi}f_{\pi}}\kappa_{5}^{'}]^{2}\;.
\end{eqnarray}
\end{subequations}

In calculations above, the normalization relations for annihilation
operators $P_{a}$, $P_{a\mu}^{*}$, $P_{0a}^{*}$, $P_{1a\mu}^{'}$ are
\begin{equation}\nonumber
   \begin{aligned}
\langle0|P_{a}|Q\bar{q}(0^{-})\rangle =\sqrt{M_{H}},&&\langle0|P^{*}_{a\mu}|Q\bar{q}(1^{-})
\rangle =\varepsilon_{\mu}\sqrt{M_{H}}\;,\\
\langle0|P^{*}_{0a}|Q\bar{q}(0^{+})\rangle
=\sqrt{M_{S}},&&\langle0|P^{'}_{1a\mu}|Q \bar{q}(1^{+})\rangle
=\varepsilon_{\mu}\sqrt{M_{S}} \;.
   \end{aligned}
  \end{equation}

\section{Numerical results}
\label{sec4}

In the numerical evaluation, the quark masses and coupling constant are adopted as
$m_{u}=2.2^{+0.6}_{-0.4}$ MeV, $m_{d}=4.7^{+0.5}_{-0.4}$ MeV, $\bar{m}=\frac{m_{u}+m_{d}}{2}=3.5_{-0.3}^{+0.7}$ MeV,
 $m_{s}=96^{+8}_{-4}$ MeV and $f_{\pi}=130.4$ MeV \cite{PDG2016}, and we get $m_{d}-m_{u}=2.5^{+0.8}_{-0.6}$ MeV by
 adding the errors in quadrature. The $\eta-\pi^{0}$ mixing angle is $\theta\simeq0.01$ \cite{Gasser1985}.

We firstly estimate $h$ by fitting experimentally measured decay widths of
the $S$ doublet mesons in Table \ref{Dmesonsdata} considering only the leading order contribution.
We obtain $h=0.50\pm0.05$ from $D^{*0}_{0}(2400)$,
$h=0.43\pm0.03$ from $D^{*\pm}_{0}(2400)$, and $h=0.71\pm0.19$ from $D^{'0}_{1}(2430)$. 
The errors come from the uncertainties of the measured masses and widths of the mesons in that doublet.
The weighted average is
\begin{equation}\label{leadh}
  h=0.44\pm0.02,
\end{equation}
with $\chi^2/2=1.78$. As is commented in Ref. \cite{PDG2016}, though acceptable, this fit is not good and we need to scale up the error by a factor of $s=\sqrt{(\chi^2/2)}=1.33$.
Thus we get finally $h=0.44\pm0.03$. However, we can see from it that it is necessary to do the further calculation
beyond the leading order. This result nicely agrees with the results from the effective
Lagrangian approach \cite{Colangelo2012}, the QCD sum rules outcome \cite{Colangelo1998} and the lattice QCD determination \cite{Becirevic}. 
With the gained $h$, at the leading order, we compute the single-pion decay widths of the $c\bar{s}$ $S$ doublet $D_{s0}^{*}(2317), D_{s1}^{'}(2460)$ and of their
bottom partners, shown separately in Table \ref{tab:table3} and Table \ref{tab:table4}. Moreover, fitted widths of the $c\bar{q}$ $S$ doublet mesons are
shown in Table \ref{widthcompare} to see the effectiveness of this weighted average approach. The errors therein are contributed by the error of $h$
and the uncertainties of their masses. It can be seen from the results that considering only the leading order is actually not enough, which proves the
necessity of doing the chiral symmetry-breaking corrections calculation beyond the leading order.

\begin{table}
\caption{\label{widthcompare} Comparison of the single-pion decay widths (all in $\mbox{MeV}$) of $c\bar{q}$ $S$ doublet mesons from
experiments $\Gamma$(exper), with only leading order contributions $\Gamma$ (leading) and the ones including
 the chiral-symmetry breaking terms $\Gamma$ (full).}
\begin{tabular}{cccc}
\hline\hline\noalign{\smallskip}
                               &  $D^{*0}_{0}(2400)$ &$D^{*\pm}_{0}(2400)$ & $D^{'0}_{1}(2430)$ \\
  \noalign{\smallskip}\hline\noalign{\smallskip}
   Mass($\mbox{MeV}$)                 & $2318\pm29$    & $2351\pm7$   &   $2427\pm26\pm25$ \\
   $\Gamma$(exper)             & $267\pm40$     &  $230\pm17$  &$384\pm^{107}_{75}\pm74$  \\
$\Gamma$(leading)           & $236\pm43$     & $275\pm21$       &$217\pm75$  \\
   $\Gamma$(full)              & $252\pm52$ & $282\pm36$      &$245\pm90$  \\
\noalign{\smallskip}\hline\hline
\end{tabular}
\end{table}

Before conducting the minimization of $\chi^2$ to extract the unknown symmetry-breaking coupling constants $\kappa_{i}$s, we need to acquire the reasonable ranges of these parameters,
assuming the corrections to be moderate and thus maintaining the convergence of the perturbation series. Specifically, following the approach of Ref. \cite{Stewart1998}, we assume that each correction
term change the leading order contribution by less than $30\%$.
Setting $h^{'}$ to be just the leading order weighted average result
$h=0.44\pm0.03$, the upper bounds of these parameters are roughly obtained:
\begin{eqnarray}\label{upper bounds}
&&|\kappa_{1}^{'}|<1.35, \quad |\kappa_{3}|<1.40,\quad
 |\kappa_{4}^{'}|<1.35,\quad |\kappa_{5}^{'}|<0.50.
\end{eqnarray}
Here, we should be aware that though the coefficients of $\kappa_{1}^{'}$, $\kappa_{3}$ and $\kappa_{4}^{'}$ are of order
$1\%$ , i.e. $4(m_{d}\pm m_{u})/{\Lambda_{\chi}f_{\pi}}\sim10^{-4}$, compared to the coefficients of $h^{'}$ and $\kappa_{5}^{'}$
for $c\bar{q}$ mesons (see Eq. (\ref{decayrates})),
they are at the same order for the $c\bar{s}$ mesons due to the relatively large value of $m_{s}$ (see Eq. (\ref{23142460decayrates})).
As a result, the upper bounds of all the $\kappa_{i}$s are of the same order.

Next, we will determine the $1/\Lambda_{\chi}$ chiral symmetry-breaking
coupling constants within their bounds using the available experimental data by $\chi^2$ fitting following the approach in Ref. \cite{Godfrey1991}.
The $\chi^{2}$ function is
\begin{eqnarray}\label{chi2}
  \chi^{2}=\sum_{i=1}^{3}\frac{(\Gamma_{theo}^{(i)}-\Gamma_{exp}^{(i)})^{2}}{(\delta\Gamma_{exp}^{(i)})^{2}},
\end{eqnarray}
where $\Gamma_{exp}^{(i)}$ and $\delta\Gamma_{exp}^{(i)}$ are the experimentally measured widths and errors of $D^{*0}_{0}(2400)$,
$D^{*\pm}_{0}(2400)$ and $D^{'0}_{1}(2430)$; $\Gamma_{theo}^{(i)}$ are the numerical values corresponding to a set of given
symmetry-breaking coupling constants. Herein, as the chiral corrections are included, the leading order coupling
constant $h$ is transferred to $h^{'}$ and should be treated as an unknown parameter now. Thus, with the transformations of the couplings
(listed in Section \ref{sec3}), in total, there are five parameters to be determined.
However, only two of them are truly free as three ones are totally constrained by three experimentally measured decay widths
in the $S$ doublet (listed in Table \ref{Dmesonsdata}). Furthermore, in the case of $c\bar{q}$ $S$ doublet mesons, coefficients of $\kappa_{3}$ and
$\kappa_{4}^{'}$ are smaller than those of $\kappa_{1}^{'}$ and much smaller (of order $1\%$)
than those of $h^{'}$ and $\kappa_{5}^{'}$. Thus considering that all the parameters are constrained
within the allowed ranges in Eq. (\ref{upper bounds}), we can infer that almost all the contributions to the decay rates of $c\bar{q}$ mesons
should be attributed to the $h^{'}$, $\kappa_{5}^{'}$ and $\kappa_{1}^{'}$ terms. Therefore the minimization of $\chi^2$
should be reliable on $h^{'}$, $\kappa_{5}^{'}$ and $\kappa_{1}^{'}$, while less reliable on
$\kappa_{3}$ and $\kappa_{4}^{'}$. However, it should be kept in mind that the $\kappa_{3}$ and $\kappa_{4}^{'}$ terms cannot be neglected,
since they are important to decay rates of the $c\bar{s}$ doublet mesons as can be seen from Eq. (\ref{23142460decayrates}).

For comparison, we do firstly the $\chi^2$ fitting of $h^{'}$ by setting all $\kappa_{i}$s to be $0$.
The result is $h^{'}=0.48\pm0.02$, which nicely agrees with the weighted average in Eq. (\ref{leadh}), but the corresponding
$\chi^{2}/2$ is pretty large, varying from $1.44.$ to $1.86$ (uncertainties originate from the choices of the masses of
particles involved within their measured errors). This also indicates that it is not enough taking into account only the leading order contributions.
Then including the chiral corrections, we carry out the minimization in its five dimensional domain ($h^{'}$, $\kappa_{1}^{'}$
$\kappa_{3}$, $\kappa_{4}^{'}$ and $\kappa_{5}^{'}$, two of which are truly free) within the bounds in Eq. (\ref{upper bounds}).
We find that the minimum of $\chi^{2}$ depends mainly on the choice of the masses of involved particles within their measured errors,
especially for $D_{0}^{*0}$, $D_{0}^{*+}$ and $D_{1}^{'}$, and also on the starting point of $h^{'}$ and $\kappa_{i}$s within their bounds slightly.
Therefore, we repeat the procedure of optimization with different sets of starting point of $\kappa_{i} s$ and of the masses
of involved particles until we are confident that we have found the absolute minimum. The final couplings are shown in
Table \ref{tab:table2}. The errors therein stem from the fact that the value of $\chi^{2}$ varies slightly and even indistinguishably around
the absolute minimum when we slightly change the values of the couplings acquired.
The corresponding $\chi^{2}/2$ is $0.66\pm0.01$, satisfying the demand $\chi^{2}/2<1$. This suggests that our optimization
is effective here and that the result is truly improved in comparison with just the leading order calculation, which are also demonstrated in the comparisons in Table \ref{widthcompare}. The main contribution to the value of $\chi^{2}$ comes from the relatively large discrepancy between the decay rates of $D_{0}^{*}$ and $D_{1}^{'}$ mesons.

\begin{table}
\caption{\label{tab:table2} Results of the six $1/\Lambda_{\chi}$ parameters by minimizing $\chi^{2}$ (only central values).}
\begin{tabular}{cccccc}
\hline\hline\noalign{\smallskip}
  $h^{'}$ & $\kappa_{1}^{'}$ & $\kappa_{2}$ &  $\kappa_{4}^{'}$ & $\kappa_{5}^{'} $ &$\chi^{2}/2$\\
\noalign{\smallskip}\hline\noalign{\smallskip}
    $0.56\pm0.01$ & $0.86\pm0.11$ & $0.52\pm0.26$  & $0.40\pm0.31$ & $0.28\pm0.03$ & $0.66\pm0.01$\\
\noalign{\smallskip}\hline\hline
\end{tabular}
\end{table}

We now turn to the estimation of the single-pion strong decay rates of
$D_{s0}^{*}(2317)$ and $D_{s1}^{'}(2460)$. With all the couplings determined,
the numerical results of their decay rates are
\begin{eqnarray}
  \Gamma(D_{s0}^{*}(2317)\rightarrow D_{s}^{+}\pi^{0}) &=& 9.2\pm2.3\; \text{KeV}, \nonumber\\
  \Gamma(D_{s1}^{'}(2460)\rightarrow D_{s}^{*+}\pi^{0}) &=& 9.0\pm2.1\;\text{KeV},
\end{eqnarray}
where the errors are from the uncertainties of their measured masses and the couplings. The results are
consistent with the experimental constraints in the second line of Table \ref{Dmesonsdata} and comparable with
other theoretical works in the literature as is demonstrated in Table \ref{tab:table3}. It
is shown that both $D_{s0}^{*}(2317)$ and $D_{s1}^{'}(2460)$ are quite
narrow and the chiral symmetry-breaking corrections are significant
compared to the leading-order ones,
the main reason of which is that $m_{s}$ is relatively large.
 In details, in the $c\bar{s}$ picture, our results are pretty close to those of
Ref. \cite{Godfrey2003,Liu2006} in the constituent
quark model, Ref. \cite{Colangelo2003} based on heavy quark symmetries and
vector meson dominance ansatz and Ref. \cite{Nielsen2006}
using the QCD sum rules in the four-quark picture,
while larger than those of Ref. \cite{Fayyazuddin2004,Wei2006,Lu2006}
using a potential model, light cone QCD sum rules and the $^{3}P_{0}$ model respectively,
but much lower than those of Ref. \cite{Ishida2004,Cheng2003}
in the covariant level-classification scheme and the four-quark picture. Moreover, in expectation, our results are much lower than those of Ref. \cite{ValeryLyubovitskij2007,Guo2014,Guo2008,Guo2013} in the molecule picture.

\begin{table}
\caption{\label{tab:table3} Strong decay rates of $D_{s0}^{*}(2317)$ to
$D_{s} \pi^{0}$ and  $D_{s1}^{'}(2460)$ to $D^{*}_{s} \pi^{0}$ (in \mbox{KeV}).}
\begin{tabular}{lcc}
\hline\hline\noalign{\smallskip}
        Approach  &  $\Gamma(D_{s0}^{*}\rightarrow D_{s}\pi^{0})$  & $\Gamma(D_{s1}^{'}\rightarrow D_{s}^{*}\pi^{0})$\\

  \noalign{\smallskip}\hline\noalign{\smallskip}
  Experiments \cite{PDG2016}       & $<3.8 \mbox{MeV}$             & $<3.6 \mbox{MeV}$ \\
  Ref. \cite{Bardeen2003}          & $21.5$              & $21.5$    \\
  Ref. \cite{Fayyazuddin2004}      & 16                  & 32       \\
  Ref. \cite{Wei2006}              & $34-44$             & $35-51$ \\
  Ref. \cite{Lu2006}               & 32                  & 35       \\
  Ref. \cite{Godfrey2003}          & $\approx10$         & $\approx10$\\
  Ref. \cite{Liu2006}              & $3.68-8.71$         & $1.86-4.42$\\
  Ref. \cite{Colangelo2003}        & $7\pm1$             & $7\pm1$  \\
  Ref. \cite{Ishida2004}           & $150\pm70$          & $150\pm70$\\
  Ref. \cite{ValeryLyubovitskij2007}& $46.7-111.9$       & $50.1-79.2$\\
  Ref. \cite{Guo2014}              & $96\pm19$           & $78\pm14$\\
  Ref. \cite{Guo2008}              & $180\pm110$         &  -         \\
  Ref. \cite{Guo2013}              & $133\pm22$          &  -         \\
  Ref. \cite{Cheng2003}            & $10-100$            &  -        \\
  Ref. \cite{Nielsen2006}           & $6\pm2$            &  -         \\
 $\Gamma$(leading)    & $5.0\pm1.0$         & $4.9\pm1.0$\\
      $\Gamma$(full)               & $9.2\pm2.3$         & $9.0\pm2.1$ \\
\noalign{\smallskip}\hline\hline
\end{tabular}
\end{table}

\begin{figure}[bp]
\centering
\begin{minipage}[t]{0.45\linewidth}
    \centering
    \includegraphics[width=5cm,height=2.5cm]{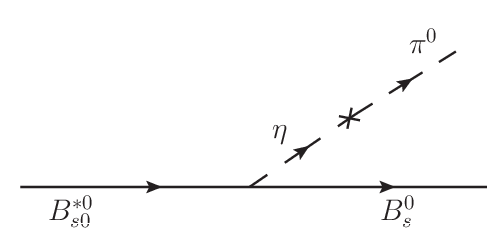}
    \parbox{3.5cm}{\small \hspace{0cm}(a) }
\end{minipage}
\hspace{2ex}   
\begin{minipage}[t]{0.45\linewidth}
    \centering
    \includegraphics[width=5cm,height=2.5cm]{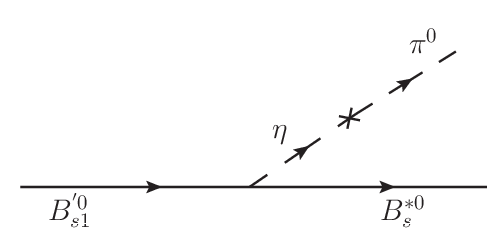}
    \parbox{3.5cm}{\small \hspace{0cm}(b) }
\end{minipage}
\caption{(a) $B_{s0}^{*0}\rightarrow B_{s}^{0}+\eta\rightarrow B_{s}^{0}+\pi^{0}$,\;
  (b) $B_{s1}^{'0}\rightarrow B_{s}^{*0}+\eta\rightarrow B_{s}^{*0}+\pi^{0}$.}
  \label{BSdecayfig}
\end{figure}

Experimentally, no candidate of the $S$
doublet of the excited heavy-light beauty mesons is observed. Nevertheless, with masses predicted by P. Colangelo
\textit{et al}. in  Ref. \cite{Colangelo2012} under the same framework,
we can obtain the single-pion decay widths of these mesons, which have
the same formulae with the corresponding charmed ones. The masses of $B_{s0}^{*}$ and $B_{s1}^{'}$ predicted in Ref. \cite{Colangelo2012}
are
\begin{eqnarray}
  M_{B_{s0}^{*0}} &=& 5706.6\pm1.2\;\mbox{MeV}, \nonumber\\
  M_{B_{s1}^{'0}} &=& 5765.6\pm1.2\;\mbox{MeV},
\end{eqnarray}
just below the $BK$ and $BK^{*}$ thresholds ($5777$ MeV and $6125$ MeV respectively) and therefore those two mesons are
expected to be very narrow, with dominant strong decays to $B_{s}\pi^{0}$ and
$B_{s}^{*}\pi^{0}$ through also the $\eta-\pi^{0}$ mixing shown in Fig. \ref{BSdecayfig}, as is recommended in the Ref. \cite{BS0BS1REFs}. Our numerical results are shown in Table \ref{tab:table4},
with the errors originating from the errors of their predicted masses and the couplings. The approaches include the HH$\chi$PT in Ref. \cite{Colangelo2012,Bardeen2003}, the light cone QCD sum rules
in Ref. \cite{Wang2008}, the improved Bethe-Salpeter method in Ref. \cite{Wang2012} and a relativistic potential model in Ref. \cite{Matsuki2012} in the $b\bar{s}$ ($b\bar{q}$ for non-strange mesons) picture,
 and effective field theories in Ref. \cite{Guo2014} and Ref. \cite{Lyubovitskij2008} in the molecule picture. It can be  seen that the results vary widely from different approaches in literatures and our results
  are very close to the results of the improved Bethe-Salpeter method in Ref. \cite{Wang2012} and comparable with the results of Refs. \cite{Colangelo2012,Bardeen2003,Guo2014,Wang2008}, while much lower that
 those of Ref. \cite{Lyubovitskij2008}
in the molecular scenario, expectedly. We can also learn that the chiral symmetry-breaking
corrections of $b\bar{q}$ are small in comparison with the leading order contributions, while those of $b\bar{s}$ are significant, which is mainly because that $m_{s}$ is relatively large compared to $m_{q}$.

\begin{table}
\caption{\label{tab:table4} Predicted masses of the $S$ doublet beauty mesons (in MeV). And strong decay widths of $B_{0}^{*}$
and $B_{1}^{'}$ (in MeV), $B_{s0}^{*}$ and $B_{s1}^{'}$ (in KeV).}
\begin{tabular}{p{1.8cm}cccc}
\hline\hline\noalign{\smallskip}
                               &$B^{*}_{0}(0^{+})$ & $B^{'}_{1}(1^{+})$  & $B^{*}_{s0}(0^{+})$ & $B^{'}_{s1}(1^{+})$ \\
  \noalign{\smallskip}\hline\noalign{\smallskip}
   Mass\cite{Colangelo2012}      & $5708.2\pm22$   & $5753.3\pm31$   & $5706.6\pm1.2$  & $5765.6\pm1.2$ \\
   $\Gamma$\cite{Colangelo2012} & $269\pm58$      & $268\pm70$      & -               & - \\
   $\Gamma$\cite{Bardeen2003}    & -               & -               & 21.5            & 21.5     \\
   $\Gamma$\cite{Guo2014}        & -               & -               & $0.8\pm0.8$     & $1.8\pm1.8$\\
   $\Gamma$\cite{Wang2008}       & -               & -               & $6.8-30.7$      & $5.7-20.7$\\
   $\Gamma$\cite{Wang2012}      & -               & -               & $13.6\pm5.6$    & $13.8\pm3.6$ \\
   $\Gamma$\cite{Matsuki2012}    & 87              & 93              & 1.6             & 1.9\\
   $\Gamma$\cite{Lyubovitskij2008}&-               & -               & $55.2-89.9$     & $57.0-94.0$\\
  $\Gamma$(leading)      & $284\pm47$      &$286\pm52$       & $6.5\pm0.1$   &$7.1\pm0.1$\\
   $\Gamma$(full)               & $313\pm53$       &$314\pm67$       & $11.6\pm1.6$    &$12.3\pm1.7$\\
\noalign{\smallskip}\hline\hline
\end{tabular}
\end{table}

In summary, we investigate the strong decays of
the exotic states $D_{s0}^{*}(2317)$ and $D_{s1}^{'}$ $(2460)$, within the
framework of HH$\chi$PT. Considering the chiral symmetry-breaking effects,
the effective heavy hadron chiral Lagrangian up to terms of next-to-leading
order in $1/\Lambda_{\chi}$ is given.
Single-pion decay widths of charmed
heavy mesons and the corresponding beauty ones in the heavy quark
spin-flavor symmetry are calculated.
Using the existing experimental data of the non-strange partners of
$D_{s0}^{*}(2317)$  and $D_{s1}^{'}(2460)$, the coupling constants are
estimated by minimizing $\chi^{2}$. Numerical analysis shows that our results
are consistent with the experimental constraints and comparable with the
other theoretical works in the literature. And the chiral symmetry-breaking
corrections of $c\bar{q}$ ($b\bar{q}$) are small in comparison with the leading
order contributions, while those of $c\bar{s}$ ($b\bar{s}$) are significant due to large mass of the strange quark.
The confirmation of such predictions is expected in the near future by
experiments at the LHCb and the hadron B factories.

\section*{Acknowledgement}
This work was supported in part by the National Natural Science Foundation
of China under Contract Nos. 11675263, 11475257 and 11475258. YLL was supported by the Chinese Scholarship Council (CSC)
and JRZ was supported by the project in NUDT for excellent youth talents.
%

\begin{thebibliography}{99}
%
%

\bibitem{exp1} BABAR Collaboration, B. Aubert {\it et al}., Phys. Rev. Lett. \textbf{90},  (2003)  242001;
 CLEO Collaboration, D. Besson {\it et al}., Phys. Rev. D \textbf{68},  (2003)  032002.
\bibitem{Godfrey1991}S. Godfrey and R. Kokoski, Phys. Rev. D \textbf{43},  (1991)  1679;
S. Godfrey and N. Isgur, Phys. Rev. D \textbf{32},  (1985)  189; M. Di. Pierro and E. Eichten,
Phys. Rev. D \textbf{64}, (2001)  114004.
\bibitem{Belle} P. Krokovny {\it et al}. (Belle Collaboration),  Phys. Rev. Lett. \textbf{91}, (2003)  262002;
   Y. Mikami {\it et al}. (Belle Collaboration),  Phys. Rev. Lett. \textbf{92},  (2004) 012002.
\bibitem{FOCUS}FOCUS Collabortation, E. W. Vaandering,  arxiv: hep-ex/0406044.
\bibitem{BABAR}BABAR Collaboration, B. Aubert {\it et al}.,  Phys. Rev. Lett. \textbf{93},  (2004) 181801.
\bibitem{Colangelo2012}P. Colangelo, F. De Fazio, F. Giannuzzi, and S. Nicotri,  Phys. Rev. D \textbf{86},  (2012) 054024.
\bibitem{Mehen2004}T. Mehen, and R. P. Springer,  Phys. Rev. D \textbf{70},  (2004) 074014;
E. Kolomeitsev and M. Lutz, Phys. Lett. B \textbf{582}, (2004)  39;
M. Q. Huang,  Phys. Rev. D \textbf{69}, (2004)  114015;
Y. B. Dai, C. S. Huang, M. Q. Huang, H. Y. Jin, C. Liu, Phys. Rev. D \textbf{58},  (1998) 094032;
L. F. Gan and M. Q. Huang,  Phys. Rev. D \textbf{82},  (2010) 054035;
A. Deandrea, G. Nardulli, A. D. Polosa,  Phys. Rev. D \textbf{68},  (2003)  097501;
Y. B. Dai, C. S. Huang, C. Liu and S. L. Zhu, Phys. Rev. D \textbf{68},  (2003) 114011;
W. Lucha and F. Schobert,  Mod. Phys. Lett. A \textbf{18},  (2003) 2837;
J. Hofmann and M.F.M. Lutz, Nucl. Phys. A \textbf{733},  (2004) 142.
\bibitem{Bardeen2003}W. A. Bardeen, E. J. Eichten and C. T. Hill,  Phys. Rev. D \textbf{68},  (2003) 054024.
\bibitem{Fayyazuddin2004}Fayyazuddin and Riazuddin, Phys. Rev. D \textbf{69},  (2004) 114008.
\bibitem{Wei2006}W. Wei, P. Z. Peng and S. L. Zhu,  Phys. Rev. D \textbf{73}, (2006)  034004.
\bibitem{Lu2006}J. Lu, X. L. Chen, W. Z. Deng, and S. L. Zhu, Phys. Rev. D \textbf{73}, (2006) 054012.
\bibitem{Godfrey2003}S. Godfrey, Phys. Lett. B \textbf{568},  (2003) 254.
\bibitem{Liu2006}X. Liu, Y. M. Yu, S. M. Zhao and X. Q. Li, Eur. Phys. J. C \textbf{47}, (2006) 445.
\bibitem{Colangelo2003}P. Colangelo and F. De. Fazio,  Phys. Lett. B \textbf{570},  (2003) 180.
\bibitem{Ishida2004}S. Ishida, M. Ishida, T. Komada, T. Maeda, M. Oda, K. Yamada, and I. Yamauchi,
AIP Conf. Proc. \textbf{717}, (2004) 716.
\bibitem{ValeryLyubovitskij2007}A. Faessler, T. Gutsche, V. E. Lyubovitskij, and Y. L. Ma, Phys. Rev. D \textbf{76}, (2007) 014005;
A. Faessler, T. Gutsche, V. E. Lyubovitskij, and Y. L. Ma, Phys. Rev. D \textbf{76}, (2007) 014008.
\bibitem{Guo2014}M. Cleven, H. W. Grie\ss hammer, F. K.  Guo, C. Hanhart  and U. G. Mei\ss ner,  arXiv:1405.2242 [hep-ph].
\bibitem{Guo2008}F. K. Guo, C. Hanhart, S. Krewald, U. G. Mei\ss ner, Phys. Lett. B \textbf{666}, (2008)  251.
\bibitem{Guo2013}L. Liu, K. Orginos, F. K. Guo, C. Hanhart, and U. G. Mei\ss ner, Phys. Rev. D \textbf{87}, (2013) 014508.
\bibitem{Barnes2003}T. Barnes, F. E. Close, H. J. Lipkin,  Phys. Rev. D \textbf{68},  (2003) 054006.
\bibitem{Cheng2003}H. Y. Cheng and W. S. Hou,  Phys. Lett. B \textbf{566},  (2003) 193.
\bibitem{Nielsen2006} M. Nielsen, Phys. Lett. B \textbf{634}, (2006) 35.
\bibitem{Browder2004}T. E. Browder, S. Pakvasa, and A. A. Petrov,  Phys. Lett. B \textbf{578},  (2004) 365.
\bibitem{Szczepaniak2003}A.P. Szczepaniak,  Phys. Lett. B \textbf{567},  (2003) 23.
\bibitem{Terasaki2003}K. Terasaki, Phys. Rev. D \textbf{68}, (2003) 011501;
J. Vijande, F. Fernandez, A. Valcarce, Phys. Rev. D \textbf{73}, (2006) 034002;
S. M. Gerasyuta, V. I. Kochkin, Phys. Rev. D \textbf{78}, (2008) 116004.
\bibitem{Mohler2013}D. Mohler, C. B. Lang, L. Leskovec, S. Prelovsek and R. M. Woloshyn, Phys. Rev. Lett.
\textbf{111}, (2013) 222001.
\bibitem{Segovia2012}J. Segovia, C. Albertus, E. Hern\'{a}ndez, F. Fern\'{a}ndez, and D. R. Entem,
Phys. Rev. D \textbf{86}, (2012) 014010.
\bibitem{chiraltheory}G. Burdman and J. F. Donoghue,  Phys. Lett. B \textbf{280},  (1992) 287;
 M. B. Wise,  Phys. Rev. D \textbf{45},  (1992) R2188;
 T. M. Yan, H. Y. Cheng, C. Y. Cheung, G. L. Lin, Y. Lin, and H. L. Yu, Phys. Rev. D \textbf{46},  (1992) 1148.
\bibitem{FOCUSD00} J. M. Link {\it et al.} (FOCUS Collaboration), Phys. Lett. B \textbf{586}, (2004) 11.
\bibitem{Belle2004} Abe K, {\it et al}. (Belle Collaboration), Phys. Rev. D. \textbf{69}, (2004) 112002.
\bibitem{PDG2016} C. Patrignani et al. (Particle Data Group), Chin. Phys. C, \textbf{40}, (2016) 100001.
\bibitem{LHCb15X} R. Aaij {\it et al}. (LHCb Collaboration), Phys. Rev. D \textbf{92}, (2015) 012012.
\bibitem{LHCb15Y} R. Aaij {\it et al}. (LHCb Collaboration), Phys. Rev. D \textbf{92}, (2015) 032002.
\bibitem{chiral-L}C. G. Boyd and B. Grinstein,  Nucl. Phys. B \textbf{442}, (1995)  205; M. Q. Huang, Y. B. Dai, C. S. Huang, Phys. Rev. D \textbf{52}, (1995)  3986; I. W. Stewart,  Nucl. Phys. B \textbf{529}, (1998)  62.
\bibitem{Fajfer2006}S.Fajfer and J. Kamenik,  Phys. Rev. D \textbf{74},  (2006) 074023.
\bibitem{McNeile2004}C. McNeile, C. Michael, and G. Thompson,  Phys. Rev. D \textbf{70},  (2004) 054501.
\bibitem{Abada2004}A. Abada, D. Be\'{c}irevi\'{c}, P. Boucaud, G. Herdoiza, J. P. Leroy,
A. Le Yaouanc, and O. P\`{e}ne, JHEP \textbf{02},  (2004) 016.
\bibitem{Stewart1998} I. W. Stewart, Nucl. Phys. B \textbf{529}, (1998) 62.
\bibitem{Gasser1985}J. Gasser, H. Leutwyler, Nucl. Phys. B \textbf{250}, (1985) 465.
\bibitem{Colangelo1998} C. K. Chow and D. Pirjol, Phys. Rev. D. \textbf{54}, (1996) 2063;
P. Colangelo and F. De Fazio, Eur. Phys. J. C \textbf{4}, (1998) 503.
\bibitem{Becirevic} D. Becirevic, E. Chang and A. L. Yaouanc, arXiv: 1203.0167 [hep-lat].
\bibitem{BS0BS1REFs} P. Colangelo and F. De Fazio, Phys. Lett. B \textbf{570}, (2003) 180, P. Colangelo, F. De Fazio, and R. Ferrandes, Mod. Phys. Lett. A \textbf{19}, (2004) 2083, P. Colangelo, F. De Fazio, and R. Ferrandes, Phys. Lett. B \textbf{634}, (2006) 235.
\bibitem{Wang2008}Z. G. Wang,  Eur. Phys. J. C \textbf{56},  (2008) 187.
\bibitem{Wang2012}Z.H. Wang, G.L. Wang, H.F. Fu, Y. Jiang,  Phys. Lett. B \textbf{706},  (2012) 389.
\bibitem{Matsuki2012}T. Matsuki and K. Seo,  Phys. Rev. D \textbf{85},  (2012) 014036.
\bibitem{Lyubovitskij2008}A. Faessler, T. Gutsche, V. E. Lyubovitskij and Y. L. Ma, Phys. Rev. D \textbf{77}, (2008) 114013.


\end{thebibliography}
%

\end{document}